\documentclass{emulateapj}

\usepackage{apjfonts}
\usepackage{amsmath}
\usepackage{amssymb}
\usepackage{natbib}
\usepackage{xspace}
\usepackage{graphicx}
\usepackage{rotating} 
\usepackage{subfigure}
\usepackage{float}

\newcommand{\err}[2]{\ensuremath{^{+#1}_{-#2}}\xspace}
\newcommand{\e}[1]{\ensuremath{^{#1}}\xspace}
\newcommand{\ten}[1]{\ensuremath{\times 10^{#1}}\xspace}

\newcommand{\nh}{$N_{\text{H}}$\xspace}

\newcommand{\pexrav}{\textsc{pexrav}\xspace}
\newcommand{\pexmon}{\textsc{pexmon}\xspace}
\newcommand{\xspec}{\textsc{xspec}\xspace}
\newcommand{\mytorus}{MYT\textsc{orus}\xspace}
\newcommand{\etal}{et al.\xspace}
\newcommand{\ka}{K$\alpha$\xspace}

\newcommand{\chidof}{$\chi^{2}/$dof\xspace}

\newcommand{\nustar}{\textsl{NuSTAR}\xspace}
\newcommand{\xte}{\textsl{RXTE}\xspace}

\newcommand{\sax}{\textsl{BeppoSAX}\xspace}
\newcommand{\suzaku}{\textsl{Suzaku}\xspace}

\newcommand{\swift}{\textsl{Swift}\xspace}
\newcommand{\xrt}{\textsl{Swift}-XRT\xspace}
\newcommand{\bat}{\textsl{Swift}-BAT\xspace}
\newcommand{\xmm}{\textsl{XMM-Newton}\xspace}
\newcommand{\chandra}{\textsl{Chandra}\xspace}
\newcommand{\fluxunits}{erg\,cm$^{-2}$\,s$^{-1}$\xspace}
\newcommand{\feunits}{photons\,cm$^{-2}$\,s$^{-1}$\xspace}
\newcommand{\crhunits}{photons\,keV$^{-1}$\,cm$^{-2}$\,s$^{-1}$\xspace}
\newcommand{\colunits}{cm$^{-2}$\xspace}

\shorttitle{The \nustar View of NGC~7582}
\shortauthors{Rivers et al.}

\begin{document}

\title{The \nustar View of Reflection and Absorption in NGC~7582} 
\author{E.~Rivers\altaffilmark{1}, M.~Balokovi\'c\altaffilmark{1}, P.\ Ar\'evalo\altaffilmark{2},  F.E. Bauer\altaffilmark{3,4,5}, S.E.~Boggs\altaffilmark{6}, W.N.~Brandt\altaffilmark{7,8,9}, M. Brightman\altaffilmark{1},  F.E.~Christensen\altaffilmark{10}, W.W.~Craig\altaffilmark{6}, P.~Gandhi\altaffilmark{11},  C.J.~Hailey\altaffilmark{12}, F.~Harrison\altaffilmark{1}, M.~Koss\altaffilmark{13}, C.~Ricci\altaffilmark{3}, D.~Stern\altaffilmark{14},  D.J.~Walton\altaffilmark{14,1}, W.W.~Zhang\altaffilmark{15}}

\altaffiltext{1}{Cahill Center for Astronomy and Astrophysics, California Institute of Technology, Pasadena, CA 91125, USA} 
\altaffiltext{2}{Instituto de F\'isica y Astronom\'ia, Facultad de Ciencias, Universidad de Valpara\'iso, Gran Bretana N 1111, Playa Ancha, Valpara\'iso, Chile} 
\altaffiltext{3}{Instituto de Astrof\'{\i}sica, Facultad de F\'{i}sica, Pontificia Universidad Cat\'{o}lica de Chile, 306, Santiago 22, Chile}
\altaffiltext{4}{Millennium Institute of Astrophysics, Vicu\~{n}a Mackenna 4860, 7820436 Macul, Santiago, Chile}
\altaffiltext{5}{Space Science Institute, 4750 Walnut Street, Suite 205, Boulder, Colorado 80301, USA} 
\altaffiltext{6}{Space Sciences Laboratory, University of California, Berkeley, CA 94720, USA}
\altaffiltext{7}{Department of Astronomy and Astrophysics, The Pennsylvania State University, University Park, PA 16802, USA}
\altaffiltext{8}{Institute for Gravitation and the Cosmos, Pennsylvania State University, University Park, PA 16802, USA}
\altaffiltext{9}{Department of Physics, 104 Davey Laboratory, Pennsylvania State University, University Park, PA 16802, USA}
\altaffiltext{10}{DTU Space, National Space Institute, Technical University of Denmark, Elektrovej 327, DK-2800 Lyngby, Denmark} 
\altaffiltext{11}{School of Physics \& Astronomy, University of Southampton, Highfield, Southampton SO17 1BJ, UK}
\altaffiltext{12}{Columbia Astrophysics Laboratory, Columbia University, New York, NY 10027, USA} 
\altaffiltext{13}{Institute for Astronomy, Department of Physics, ETH Zurich, Wolfgang-Pauli-Strasse 27, CH-8093 Zurich, Switzerland}
\altaffiltext{14}{Jet Propulsion Laboratory, California Institute of Technology, Pasadena, CA 91109, USA} 
\altaffiltext{15}{NASA Goddard Space Flight Center, Greenbelt, MD 20771, USA} 

\email{Contact: erivers@caltech.edu}

\begin{abstract}

NGC~7582 is a well-studied X-ray bright Seyfert 2 with moderately heavy ($N_{\text{H}}\sim10^{23}-10^{24}$~cm$^{-2}$), highly variable absorption and strong reflection spectral features.
The spectral shape changed around the year 2000, dropping in observed flux and becoming much more highly absorbed.
Two scenarios have been put forth to explain this spectral change: 1) the central X-ray source partially ``shut off'' around this time, decreasing in intrinsic 
luminosity, with a delayed decrease in reflection features due to the light-crossing time of the Compton-thick material or 2) the source became
more heavily obscured, with only a portion of the power law continuum leaking through.
\nustar observed NGC~7582 twice in 2012, two weeks apart, in order to quantify the reflection using high-quality data above 10 keV.
We find that the most plausible scenario is that NGC~7582 has recently become 
more heavily absorbed by a patchy torus with a covering fraction of $\sim\,80-90\%$ and an equatorial column density of 
$\sim 3 \times10^{24}$~cm$^{-2}$.  We find the need for an additional highly variable full-covering absorber with 
$N_{\text{H}}= 4-6 \times10^{23}$~cm$^{-2}$ in the line of sight, possibly associated with a hidden broad line region.

\end{abstract}

\keywords{X-rays: galaxies -- Galaxies: active -- Galaxies: Individual: NGC 7582}

\section{Introduction}

NGC~7582 is an X-ray bright, nearby ($z=0.00525$, $D=22.0$\,Mpc), Seyfert 2 which has recently undergone significant spectral evolution.
Spectral changes in active galactic nuclei (AGN) are indicative of a dynamic environment and understanding these spectral 
variations can help place constraints on the circumnuclear material in AGN.
Early X-ray observations of NGC~7582 by \textsl{Einstein} (Maccacaro \& Perola 1981), 
\textsl{EXOSAT} (Turner \& Pounds 1989), \textsl{Ginga} (Warwick \etal 1993), 
\textsl{ASCA} (Schachter \etal 1998), and \textsl{BeppoSAX} (Turner \etal 2000, hereafter T00) revealed a very flat observed spectrum, 
indicative of high levels of absorption and/or strong reflection.  T00 obtained one of the earliest high-quality broadband spectra of this source and concluded that the 
spectral shape was due to a full-covering layer of absorption with \nh $\sim$ 10\e{23}~\colunits and a second layer of partial-covering absorption with 
\nh $\sim$ 10\e{24}~\colunits covering $\sim$60\% of the continuum power law.

Bianchi \etal (2007) analyzed \textsl{Hubble} and \chandra data, confirming that the soft X-ray 
spectrum below 2 keV is dominated by extended emission from highly ionized diffuse 
gas in the center kiloparsec of the galaxy.  They also found that a dust lane in the host galaxy 
covers the central AGN, contributing to the obscuring material in the line of sight ($\sim 10^{22}$~\colunits).  
This is consistent with the deep silicate absorption feature seen in mid-infrared observations of this source (Asmus \etal 2014).

\begin{figure}
      \includegraphics[width=0.45\textwidth]{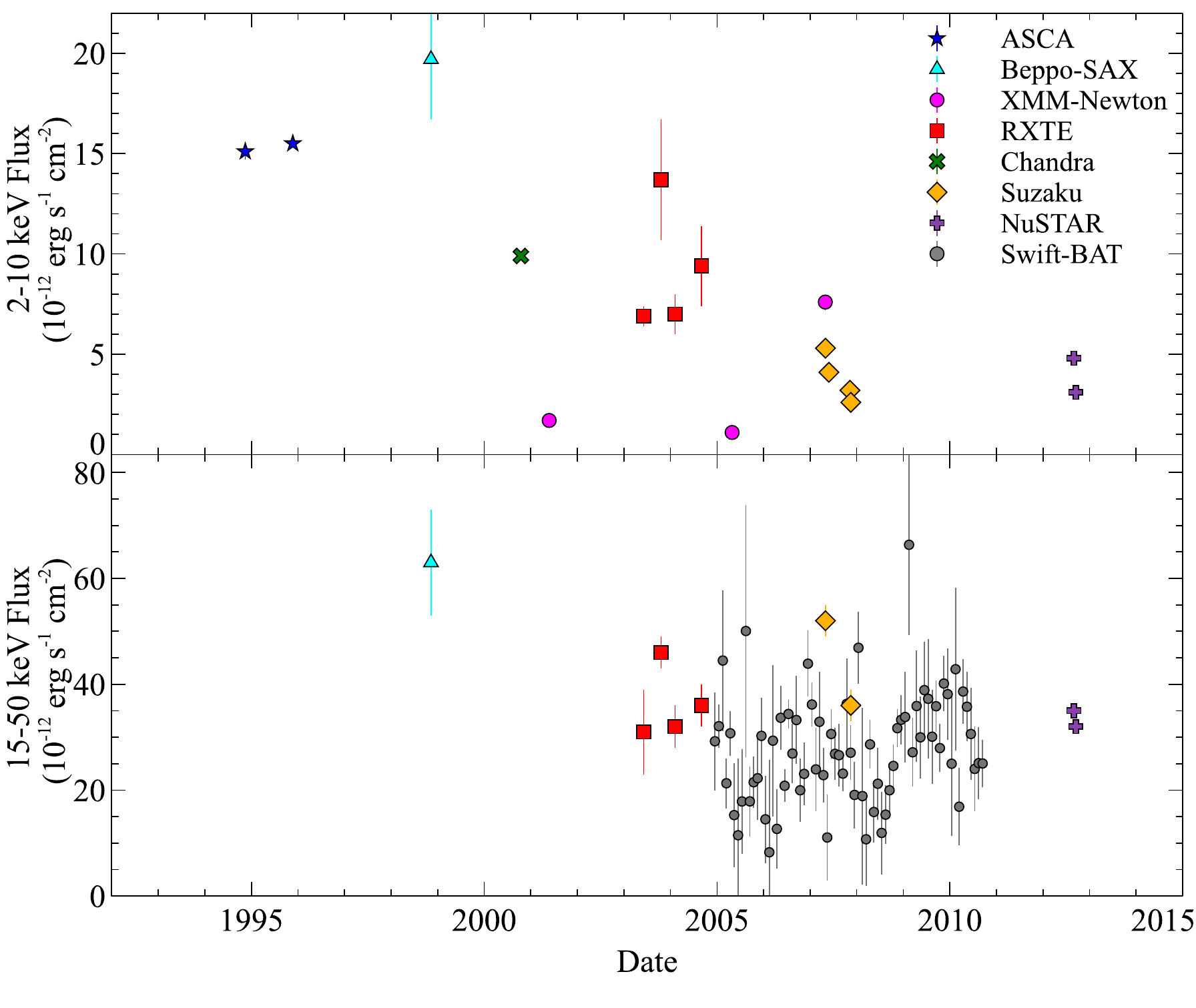}
 \caption{Top: historical 2--10 keV observed fluxes of NGC~7582 from \textsl{ASCA} (T00), \sax (T00), \xmm (P07/B09), \chandra (Bianchi \etal 2007), \xte (this work), \suzaku (B09), and \nustar (this work).  Bottom: historical 15--50 keV observed fluxes from \sax (T00), \xte (this work), \bat (Baumgartner \etal 2013), \suzaku (B09), and \nustar (this work).}
  \label{figlc}
\end{figure}

In 1998, broad H lines appeared in the optical spectrum of NGC~7582, 
followed by a slow decay of the broad lines over the course of $\sim$6 months (Aretxaga \etal 1999).
Possible explanations for this change to a Seyfert 1 type spectrum include starburst supernova activity, a tidal disruption 
event in the nucleus, or a ``changing look'' event where the obscuring material blocking the broad lines moved temporarily out of the line of sight.  
If this latter explanation were the case we might have expected a change in the absorbing material to be detectable in the X-ray spectrum of the source.
The \sax observation occurred near the end of this period and did not appear to show any significant changes in the X-ray behavior compared to previous observations,
although follow-up X-ray observations after 1998 show lower observed fluxes than pre-1998 and more spectral curvature (see Figure \ref{figlc}).
 
Piconcelli \etal (2007, hereafter P07) analyzed \xmm data from 2001 and 2005 and found that the source had undergone significant changes, 
with its observed 2--10 keV flux falling from 1.2\ten{-11} \fluxunits (from their reanalysis of the \sax observation) in 1998 to 2\ten{-12} \fluxunits in 2001, with a similar value in 2005.  
Additionally, the spectrum looked far less like a power-law in the later data, either due to increased obscuration or an increased amount of reflection compared to the continuum power law.  
P07 concluded that the spectrum was most accurately described by a high column density ($\sim10^{23}$ \colunits) absorbed continuum 
plus a ``pure reflection'' component with a full-covering Compton-thin ($\sim 10^{22}$~\colunits) absorber covering both.
They attributed the lower column density absorber to the dust lane and the thicker absorber to clumpy material near the central black hole.
P07  explained the strength of the reflection component as a ``shut-off'' of the central source after 1998, 
where it decreased in average intrinsic luminosity with a delayed decrease in reflection strength due to the light-crossing time to the Compton-thick material. 
Given the implied distance to the Compton-thick material ($\gtrsim$\,1~pc), they reasoned that this is most likely material in the infrared torus.

Bianchi \etal (2009, hereafter B09) used this same model to describe four \suzaku observations from 2007 and an additional \xmm observation, also from 2007.
They found dramatic spectral variability on timescales from <1 day to 7 months due to variable absorption of the continuum power law which they attributed to 
hidden broad line region (BLR) clouds.
They were able to keep the photon index and reflection normalization frozen between the five observations,
though due to the low signal-to-noise ratio in the PIN and the spectral coverage gap between the XIS and PIN (10--16 keV), 
they were unable to reliably test for variations in the reflection strength across the observations.
B09 did not find the need for a high reflection fraction in their \suzaku data,
instead concluding that all the spectral variability could be explained by the variable obscuration instead of large changes in the AGN intrinsic luminosity.
B09 also reanalyzed previous \xmm data and found that while the reflection fraction was high, the normalization of the reflection hump was consistent with being constant over time.

The question remains whether the reflection will eventually decrease in response to the drop in overall flux after 1998, 
as suggested by P07, or whether that decrease was in fact due to increased absorption in the line of sight, as modeled by B09, 
in which case we would not expect the reflection strength to decrease.
\nustar observed NGC~7582 twice in 2012 two weeks apart, obtaining high quality spectra from 3--78 keV.  
We also obtained a \xrt observation quasi-simultaneous to the first observation, expanding our range down to 0.2 keV.
We analyzed these spectra and past observations to attempt to disentangle the reflection and absorption model components.
This paper is structured in the following way: Section 2 contains details of the observations and data reduction, Section 3 describes the spectral analysis, and 
Section 4 discusses our results and conclusions.  
We have also included analysis of time resolved \xte data in an appendix to show long term trends.



\begin{deluxetable}{lcc}
   \tablecaption{\nustar Observation Details \label{tabobs}}
   \tablecolumns{3}
   \startdata
\hline
\hline\\[-1mm]
Observation  				&	1	&	2		\\[1mm]
\hline\\[-1mm]
Observation Date (UT)			&	2012 Aug 31	&	2012 Sep 15	\\[1mm]
\nustar ObsID				& 	60061318002	&	60061318004	\\[1mm]
FPMA Net Exposure (ks)		& 	16.5			&	14.6		\\[1mm]
FPMB Net Exposure (ks)		&  	16.4			&	14.6		\\[1mm]
\hline \\[-1mm]
Observation Date (UT)		&	2012 Sep 01	&\\[1mm]
\swift ObsID				&	32534001		& \\[1mm]
\swift Net Exposure (ks)		& 	4.0			& \\[-1cm]
\enddata
\tablecomments{Details of the \nustar and \swift observations.}
\vspace{1cm}
\end{deluxetable}

\begin{figure}
      \includegraphics[width=0.45\textwidth]{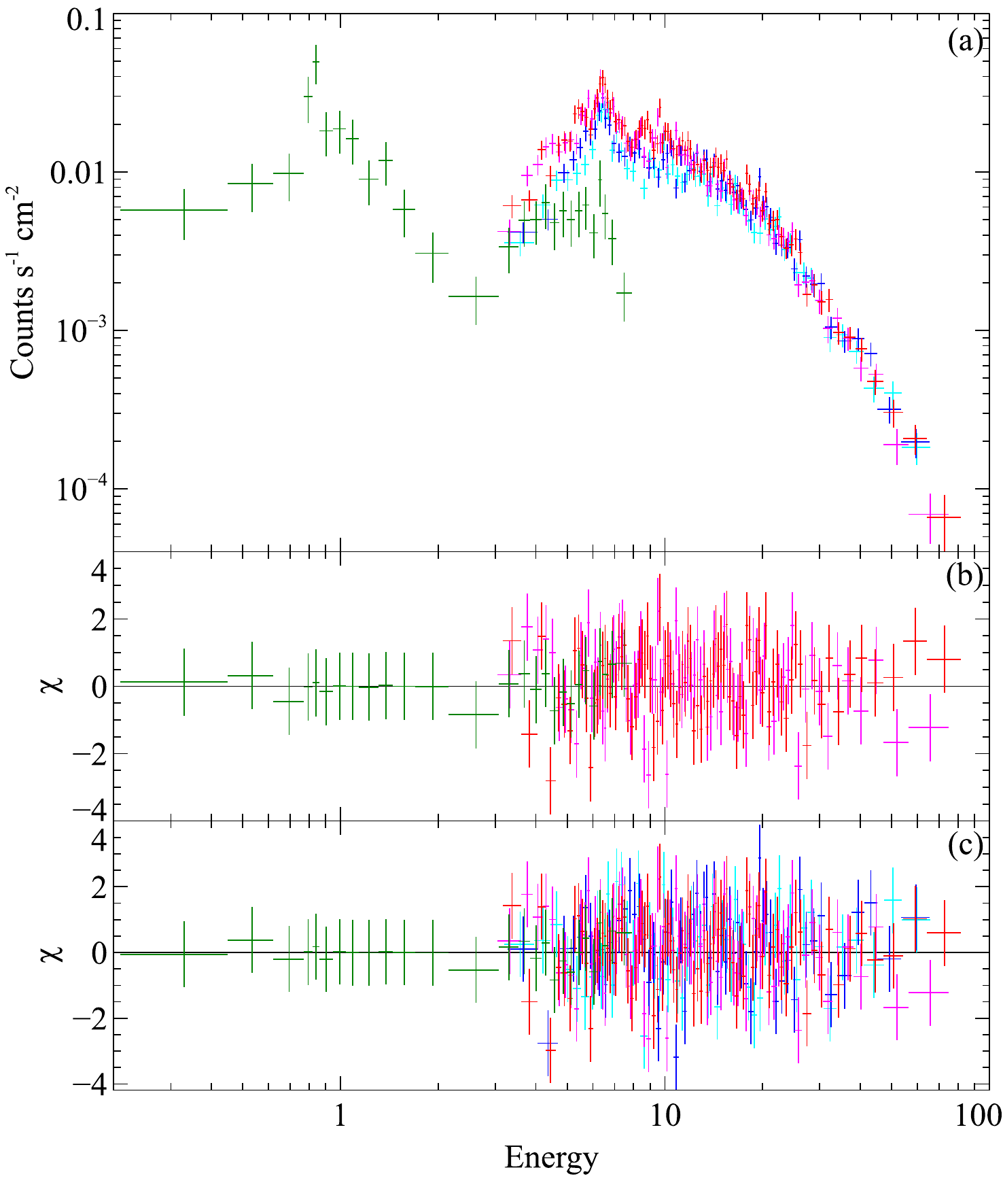}
 \caption{Spectrum and ratio plots for the \pexrav model fits.  \xrt data are shown in green, \nustar observation 1 data are shown in red and magenta (FPMA and FPMB), and \nustar observation 2 data are shown in dark blue and light blue (FPMA and FPMB).  a) \nustar and \xrt data from observations 1 and 2.  b) Data/model ratios to \pexrav model best fit of observation 1 only.  c) Data/model ratios to \pexrav model best fit of both observations simultaneously.}
  \label{figtrad}
\end{figure}

\begin{figure}
   \includegraphics[width=0.45\textwidth]{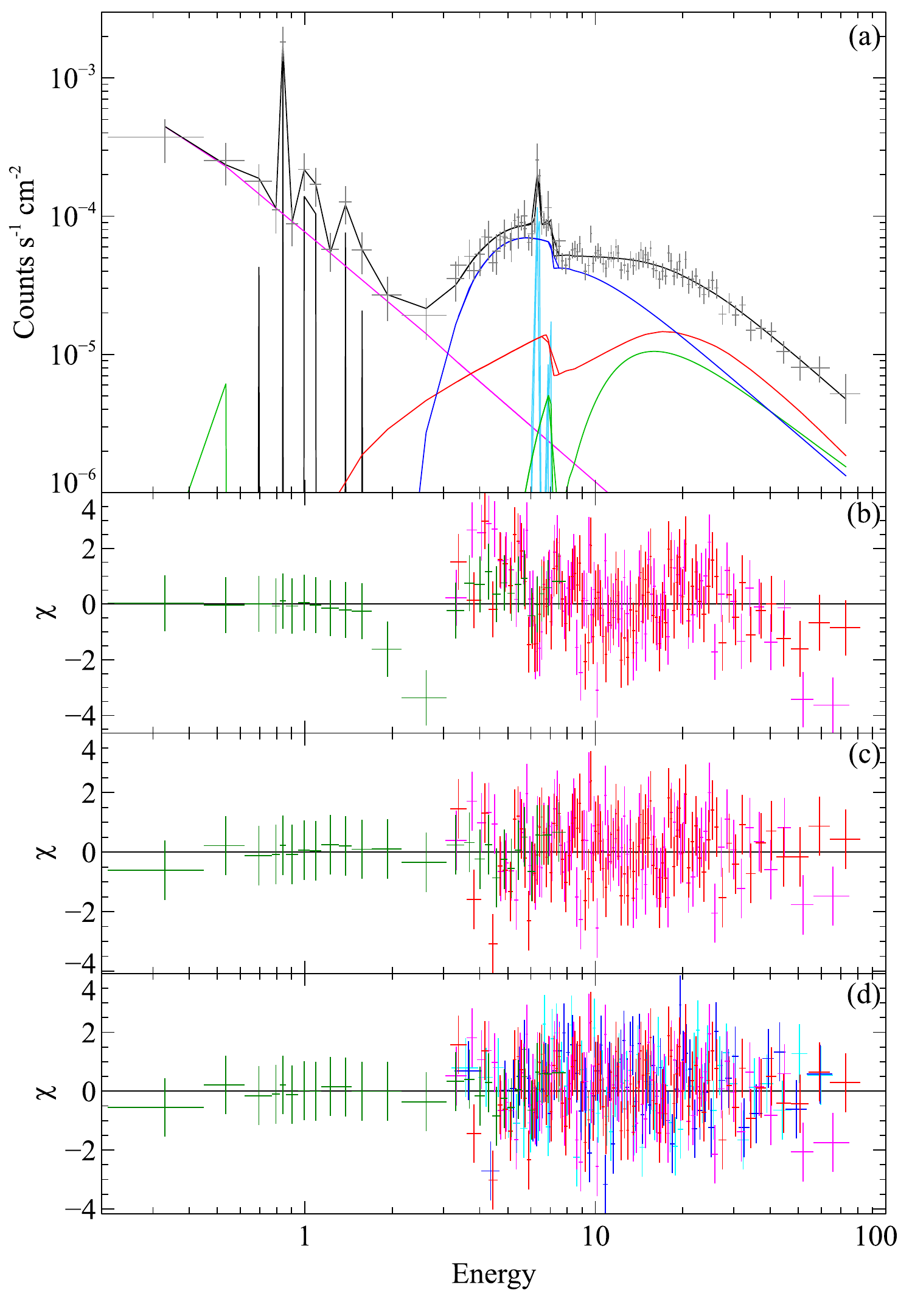}
 \caption{Results of fitting the \mytorus model (see Section 3.1). a) Best fit model (black) along with the model components: primary absorbed continuum (green), reflection hump (red), Fe line complex (cyan), leaked power law with full covering absorber (blue), soft power law (magenta) and soft emission lines (black).  Unfolded \nustar and \xrt data from observation 1 are shown in gray.  b) Data$-$model residuals to \mytorus recommended basic implementation (no leaked emission, recommended parameters tied) of observation 1 only.  c)  Data$-$model residuals to best fit \mytorus model (with leaked emission) of observation 1 only.  d) Data$-$model residuals to best fit \mytorus model of both observations simultaneously.   \xrt data are shown in green, \nustar observation 1 data are shown in red and magenta (FPMA and FPMB), and \nustar observation 2 data are shown in dark blue and light blue (FPMA and FPMB)}
  \label{figmyt}
\end{figure}


\begin{figure}
   \includegraphics[width=0.49\textwidth]{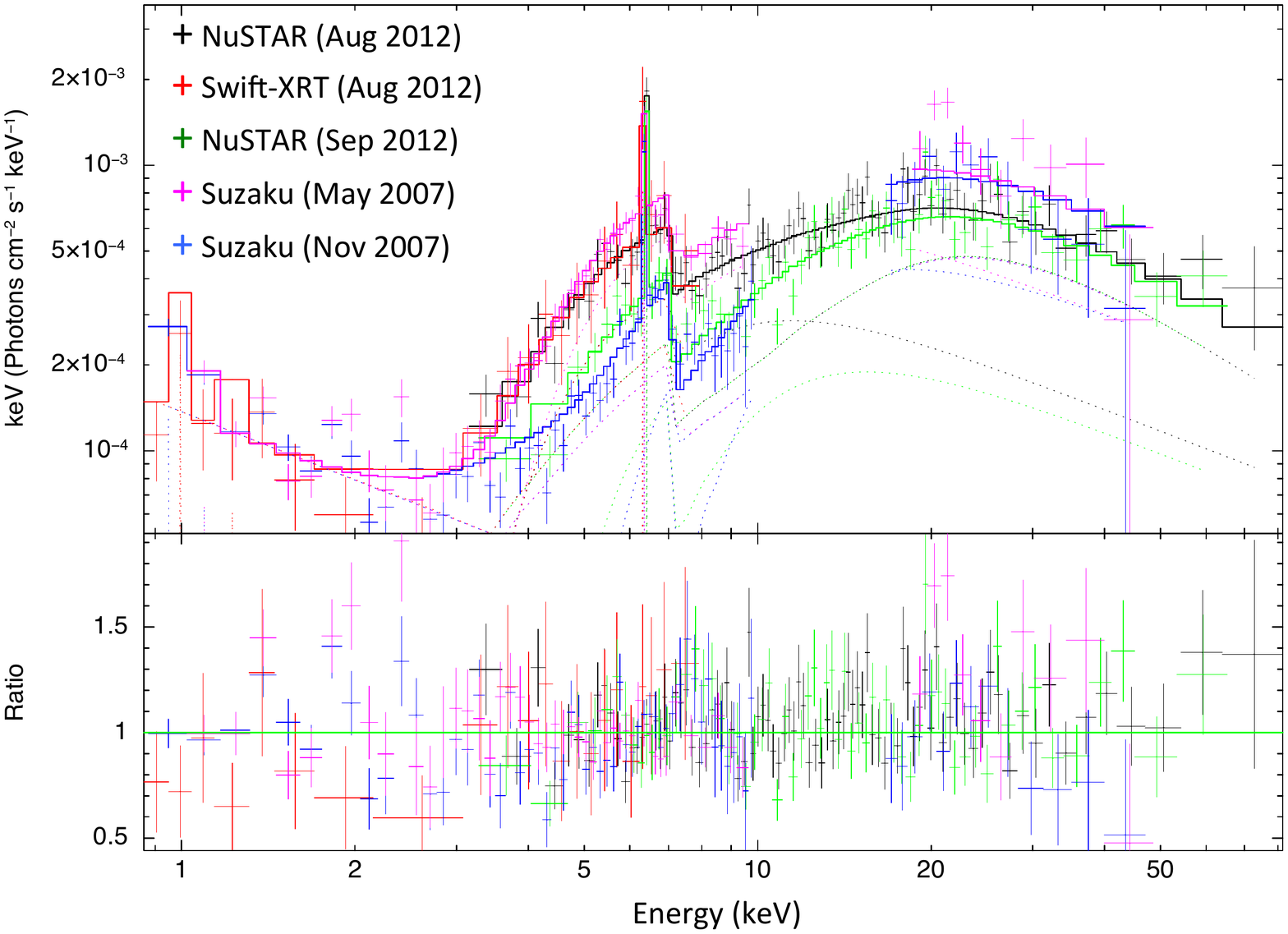}
 \caption{Unfolded data with model components and data/model ratios for simultaneous fitting (using the \pexrav model) of both \nustar observations and the first and last \suzaku observations (showing only FPMA+XRT and XIS-FI+PIN for clarity).  A good fit was found leaving free only \nh (varies by up to a factor of 5), continuum power law normalization (varies by up to a factor of 2), and a cross calibration constant between \nustar and \suzaku ($\sim$0.85).  \pexrav and  Fe line normalizations and $\Gamma$ were left tied and seem to be consistent.}
  \label{figall}
\end{figure}


\section{Observations and Data Reduction}\label{sec:analysis}

NGC~7582 was observed by \nustar on 2012 August 31 and two weeks later on 2012 September 15 for 16.5~ks and 14.6~ks, respectively. 
The observations were performed as part of the \nustar snapshot survey of AGN detected by the \bat instrument (Tueller \etal 2010, Balokovi\'c \etal 2014) 
and followed up with a 4.0~ks \xrt observation on 2012 September 1. For the remainder of this paper, we consider the \xrt observation as 
simultaneous to the first \nustar observation. Simultaneous observations with \swift and \nustar allow for overlapping spectral coverage 
where the \nustar sensitivity drops off, resulting in a broadband spectrum from approximately 0.5 to 78~keV. Table \ref{tabobs} shows a log of the observations.

\subsection{\nustar Reduction}

We reduced data from both \nustar focal plane modules, FPMA and FPMB (Harrison \etal 2013), using the standard pipeline in the \nustar 
Data Analysis Software (NUSTARDAS) version 1.4.1, distributed with HEASOFT version 6.16. Instrumental responses were taken from the 
\nustar calibration database (CALDB) version 20150316. The raw event files were cleaned and filtered for South Atlantic Anomaly (SAA) 
passages using the \texttt{nupipeline} script. The source was well detected in the entire \nustar bandpass.
The cleaned events were further processed using the \texttt{nuproducts} script, producing light curves, 
spectra and response files. For both modules we extracted spectra and light curves from a circular extraction region with a $75\arcsec$ radius 
centered on the point source. This choice maximizes the signal-to-noise ratio above 25~keV and leaves ample detector area free for background 
sampling in short, $\simeq$20~ks \nustar observations (Balokovi\'c \etal in prep). Background was extracted from polygonal regions that cover the same 
detector as the source in each focal plane module. We do not use \nustar data below 3~keV or above 78~keV. We grouped all \nustar 
spectra with a minimum of 50 counts per bin before background subtraction.

\subsection{\swift Reduction}

The \xrt observation was performed in the Photon Counting mode (Hill \etal 2004, Burrows \etal 2005). 
The data were reduced using the task \texttt{xrtpipeline} (version 0.12.6), which is a part of the XRT Data Analysis Software (XRTDAS) within HEASOFT. 
The spectrum was extracted from a circular region 20$\arcsec$ in radius centered on the point source. 
The background was extracted from a large annular source-free region encircling the source region. 
We used the response file \texttt{swxpc0to12s6\_20010101v014.rmf} from the \xrt calibration database, while the auxiliary response file was generated using the task \texttt{xrtmkarf}.  
Due to low count statistics, the \xrt spectrum is binned to a minimum of 10 photons per bin.


\begin{deluxetable*}{lcccccccr}
   \tablecaption{Broadband Model Parameters \label{tabpar}}
   \tablecolumns{9}
   \startdata
\hline
\hline\\[-1mm]
\pexrav & Observed		& Unabsorbed			 	& Photon  	&	PL		& 	Column  		& 	Fe \ka	& \pexrav 			&	$\chi^2$/dof \\[1mm]
Model  & F$_{2-10}$\tablenotemark{A}	& Power Law  	& Index 		&	Norm\tablenotemark{B}	& 	Density \tablenotemark{C}		& Line Norm\tablenotemark{D} &	Norm\tablenotemark{E}	 \\[1mm]
	   &  		& F$_{2-10}$\tablenotemark{A} 		& ($\Gamma$) 	&  &  (\nh) & 	&	 \\[1mm]
\hline\\
\nustar 1	& 4.8$\,\pm\,$0.1	 &	9.2$\,\pm\,$1.0	&	1.78$\,\pm\,$0.07	&	2.6$\,\pm\,$0.2	&	24\err{3}{7}	&	2.1$\,\pm\,$0.4		&	10$\,\pm\,$3	&	353/312	\\[1mm]
\nustar 2	& 3.2$\,\pm\,$0.1 &	7.2$\,\pm\,$1.6	&					&	2.1$\,\pm\,$0.2	&	56\err{10}{20}	&			\\[1mm]
\hline
\hline\\[-1mm]
\mytorus	 & Observed		& Unabsorbed 			& Photon  				& Absorbed PL			&	Leaked PL  		& Full Covering  			&  Torus Column 		&  $\chi^2$/dof \\[1mm]
Model	& F$_{2-10}$\tablenotemark{A}	 & Power Law  	& Index 				& Norm\tablenotemark{B}				&	Norm\tablenotemark{B}			& Column Density\tablenotemark{C}  		& Density\tablenotemark{C}			&  \\[1mm]
		& &  F$_{2-10}$\tablenotemark{A}			& ($\Gamma$) 			&		&	 		& (\nh) & (\nh)	&    \\[1mm]
\hline\\
\nustar 1	& 4.8$\,\pm\,$0.1 &	60$\,\pm\,$20		&	1.82$\,\pm\,$0.07	&	13$\,\pm\,$4	&	3.3$\,\pm\,$0.6	&	26\err{4}{7}		&	365$\,\pm\,$45		&	355/312\\[1mm]
\nustar 2	& 3.2$\,\pm\,$0.1 &			&					&	15$\,\pm\,$6	&	2.1$\,\pm\,$0.5	&	36\err{8}{13}		&		&	\\[1mm]
\enddata
\tablecomments{Model parameters to joint simultaneous fitting of \nustar/\xrt spectral data with the \pexrav and \mytorus models.  
Note we have fixed the inclination angle to 65\degr for both models.}
\tablenotetext{A}{Flux is in units of $10^{-12}$~\fluxunits.}
\tablenotetext{B}{Power law norm is in units of $10^{-3}$~counts s\e{-1} keV\e{-1} at 1 keV.}
\tablenotetext{C}{Column densities are in units of $10^{22}$\,cm$^{-2}$}
\tablenotetext{D}{Fe line flux is in units of $10^{-5}$~\feunits.}
\tablenotetext{E}{\pexrav Norm is in units of $10^{-3}$~\crhunits.}
\end{deluxetable*}

\section{Spectral Analysis}

All spectral fitting was done in \xspec v.12.8.0 (Arnaud 1996) using the solar abundances of Anders \& Grevesse (1989) and cross-sections from Verner \etal (1996). 
Uncertainties are listed at the 90\% confidence level ($\Delta \chi^2$ = 2.71 for one interesting parameter).
We included a constant offset for each instrument as a free parameter (values falling between 0.98 and 1.02) to account for known cross-calibration uncertainties (Madsen \etal 2015)
and included a Galactic absorption column of 1.33 $\times 10^{20}$ cm\e{-2} in all models (Kalberla \etal 2005).
NGC~7582 has a measured galactic inclination angle of 68\degr (see, e.g., Paturel \etal 2003) and a redshift of $z=0.00525$. 

We started with observation 1, fitting the \nustar FPMA and FPMB spectra in the 3--78 keV range simultaneously with the \xrt spectrum in the 0.3--8 keV range.
We first tried to replicate the model used by P07 and B09.  This model consists of a nuclear continuum 
absorbed by a column of $\sim\,10^{23}$~\colunits with a Gaussian Fe \ka line and a Compton reflection hump modeled by \pexrav (Magdziarz \& Zdziarski 1995), 
all which are absorbed by a dust lane in the host galaxy with the column frozen at the average value found by B09 of $4 \times 10^{22}$~\colunits
due to degeneracy with the higher column absorber.

In order to model the diffuse soft X-ray emission, the model also includes a scattered 
power law ($\Gamma$ tied to that of the continuum power law) and five distinct narrow emission lines which we have matched to the most prominent \xmm RGS and EPIC lines from P07:
0.65 keV (O \textsc{viii}),  0.85 keV (Fe  \textsc{xvii}), 1.05 keV (Fe \textsc{xxi}), 1.35 keV (Mg  \textsc{xi}) and 1.70 keV (Si \textsc{xiii}).

The \xspec notation used to describe this model is:
\begin{center}
\textsc{zphabs(zphabs\,$\times$\,power law + zgauss\,(Fe\,\ka) + pexrav) + soft\,power\,law + zGauss[$\times$5]}
\end{center}

This model fit our observation 1 data well, with \chidof$=188/179$.  Next we added the observation 2 data and investigated which parameters could be left 
tied between the two observations and which needed to be free.  The soft X-ray components were tied in all fits due to the lack of XRT data for observation 2, and
we do not expect these components to vary on such short timescales since they arise largely from extended material in the vicinity of the nucleus as discovered by \chandra (Bianchi \etal 2007).
We found that the photon index, Compton reflection strength, and Fe \ka line parameters were stable between the observations and could be tied.
The only parameters which showed a large change between the observations were the column density and the normalization of the power law (see Table \ref{tabpar}).
Data and data/model ratio plots for this model are shown in Figure \ref{figtrad}.


Given the quality of our high-energy data, we also attempted more sophisticated physical modeling that has recently been developed.  
To begin, we tried simply replacing \pexrav and the Fe \ka Gaussian line with 
the \pexmon model, which includes expected emission lines in the Fe K bandpass modeled self-consistently with the Compton reflection hump for a disk geometry.  
However, this model provided a worse fit unless the Fe abundance was allowed to be lower than 1 ($\sim\,0.6$). 
This is consistent with the equivalent width of the Fe line compared to the \pexrav component of $\sim0.5$ keV (1 keV is expected, see, e.g., B09 and Matt \etal 1991).

For both the \pexrav and \pexmon models, the Compton reflection hump is much stronger than expected, given the strength of the continuum.
For a flat disk geometry the expected \pexrav normalization is equal to the power law normalization when $R$ is set to 1, as in our model.
For other geometries such as a warped disk or a torus this value could be somewhat higher.  However, we measure a normalization that is 
four times that of the power law (equivalent to an $R$ value of 4.3), much higher than expected.
This was noted by P07 who theorized that the source may have dimmed substantially between 1998 and 2001 and that there has been a delay in the drop of the 
reflection strength, i.e., that the reflection material is very far from the source. 

\subsection{Torus Modeling}

An alternate explanation of the discrepancy is a hidden nucleus, 
where the Compton-thick torus actually intersects the line of sight to the nucleus, contributing obscuration as well as reflection.
We can test this scenario using the \mytorus model (Murphy \& Yaqoob 2009) in \xspec.
The model has three components (the absorbed zeroeth-order continuum, the Compton reflection hump, and the Fe \ka line) 
with the viewing angle frozen at 65\degr, an opening angle of the torus of 60\degr, and equatorial column density free to vary (although it is tied between the three components).  
Using the recommended fully-coupled model plus soft extended emission as modeled before (power law plus emission lines) 
could not provide an acceptable fit (\chidof$=300/185$ with $\Gamma \sim 1.4$) due to extra emission in the 3--15 keV range.  
We therefore included the same absorbed power law which in the previous model was considered the primary continuum:  

\begin{center}
\textsc{zphabs(zphabs\,$\times$\,leaked power law + MYTZ\,(0th order power law) + MYTS\,(Compton reflection)+ MYTL\,(Fe\,\ka)) + soft\,power\,law + zGauss[$\times$5]}
\end{center}

This provided a good fit to observation 1 with \chidof$=188/182$.
Data and data/model ratio plots for this model are shown in Figure \ref{figmyt}.
The relative strength of this power law compared to the inferred strength of the heavily absorbed/reflected zeroth order continuum 
is quite high to be scattered continuum, $\sim 20\%$ as compared to a more typical $<10\%$ (e.g., Noguchi \etal 2010).  
It is more likely that it is leaked emission due to a patchy absorber with a covering fraction of $f \sim 80\%$.
This power law is still absorbed by a variable $10^{23}$ \colunits column in the line of sight which may be part of the hidden BLR or the patchy torus.

Including the data from observation 2, we obtain a good fit (\chidof$=355/312$) with all parameters tied save for the leaked power law normalization and absorption column.
Using this model we find that observation 2 has a covering fraction of $f\sim90\%$ and an increase in the column density from 2.6 to 3.6 $\times 10^{23}$~\colunits in the line of sight.

In place of the \mytorus model with the extra leaked power law component, we also tried a decoupled \mytorus model.
This models clumpiness of the obscuring and reflecting medium, allowing the material in our line of sight to have a different density than the systemic average.
To do this we used the recommended full model as above, but untied the column density and the normalization between the \mytorus direct and scattered components 
(and between the observations), although degeneracy led us to tie \nh for the line component to the scattered component.
This model provided an acceptable fit (\chidof$=351/300$), not quite as good as with the leaked power law model.
The best fit \nh values were 5\ten{23} \colunits in the line of sight and 3\ten{24} \colunits out of the line of sight.
Additionally, this model led to a scattered component that was six times higher than expected and a line component that was three times higher than expected, even with the higher column.
This could potentially be explained by variation in the source power, except that we know our 20--50 keV flux is comparable to the average (see Figure 1, lower panel).
Simulating Compton scattering from clumpy material we also tried using two scattering components with inclination angles of 0\degr and 90\degr.  
This led to a good fit but with the same issue of unexpectedly high reflecting fractions.

We also applied the alternative \textsc{bntorus} model from Brightman \& Nandra (2011) including an absorbed leaked power law component as with the \mytorus model.
We found very similar absorption and continuum parameters to the \mytorus model with
an equatorial torus column density of 2.4$\times 10^{24}$~\colunits and an incident power law normalization of 1.3\ten{-2} counts s\e{-1} keV\e{-1} at 1 keV.
We measured a viewing angle of  $65\degr \pm 5\degr$ and an opening angle of $60\degr \pm 5\degr$.  
Note that for both toroidal models, the equatorial column density and therefore the reflection spectrum strength, 
is heavily influenced by the absorption column which may be variable in this source given the variability of the lower column density absorber.


\begin{figure*}
  \plottwo{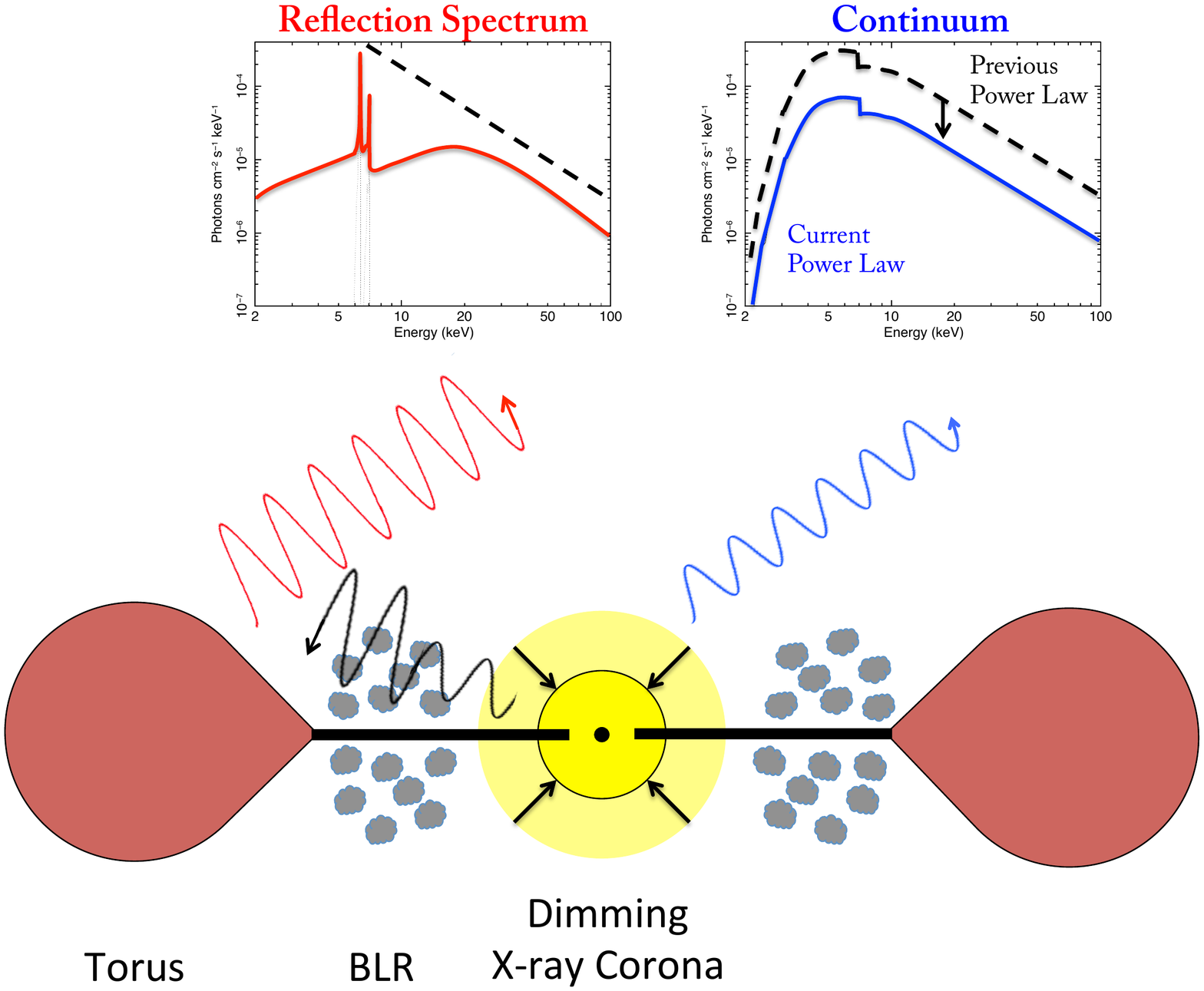}{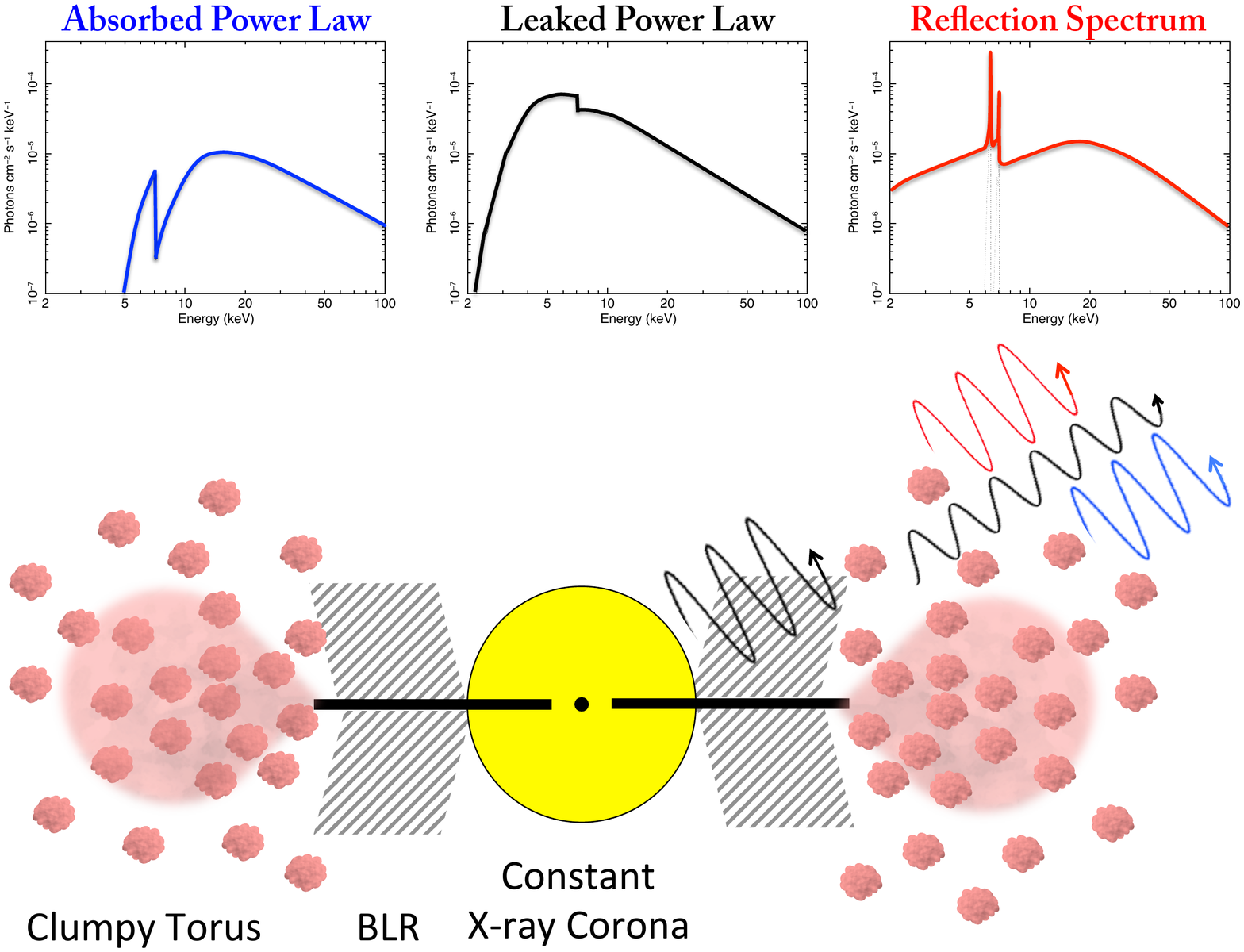}
 \caption{Illustrations of the shut-off scenario (left) and the hidden nucleus/patchy torus scenario (right).  In the shut-off scenario, the source dimmed considerably circa 2000, but the reflected spectrum has not yet decreased.  The expected time delay would be equal to the light crossing time of the torus.  In the patchy torus scenario the source contains a hidden nucleus obscured by the Compton-thick torus clouds, a reflection component, and a leaked component, either due to very small clumps not fully obscuring the central source, or to very fast moving clumps which do not obscure the source all the time.  Note that for the shut-off scenario the X-ray corona is dimming, not necessarily decreasing in size.}
  \label{figschema}
\end{figure*}


\begin{deluxetable}{lccccc}
   \tablecaption{Historical Absorption/Reflection Parameters \label{tabhist}}
   \tablecolumns{6}
   \startdata
\hline
\hline\\[-1mm]
Date		&	Observatory	&	Observed	&	\nh		&	Fe \ka 	&	\pexrav \\[1mm]
&&	F$_{2-10}$\tablenotemark{A}	&	($10^{22}$~\colunits)	&	Norm\tablenotemark{B}	&	Norm\tablenotemark{C} \\[1mm]
\hline\\
1978 Dec 		&	\textsl{Einstein}		&	64	&	8\err{2}{1}	\\[1mm]
1979 May 		&	\textsl{Einstein}		&	36	&	10\err{0.6}{0.3}	\\[1mm]
1979 Jun 	 	&	\textsl{Einstein}		&	24	&	13\err{7}{3}	\\[1mm]
1979 Nov  	&	\textsl{Einstein}		&	52	&	20\err{5}{9}	\\[1mm]
1980 May  	&	\textsl{Einstein}		&	34	&	5\err{4}{3}	\\[1mm]
1984 Jun 	 	&	\textsl{EXOSAT}	&	16	&	15\err{17}{9}	\\[1mm]
1988 Oct   	&	\textsl{Ginga}		&	27	&	48\err{25}{15}	\\[1mm]
1994 Nov  	&	\textsl{ASCA} 		&	15$\pm$4	&	8$\pm$1	&		&		\\[1mm]
1996 Nov 		&	\textsl{ASCA} 		&	16$\pm$2	&	12$\pm$1	&		&		\\[1mm]
1998	 Nov 		&	\sax		&	19.7$\pm$0.3	&	14$\pm$1	&	7.0$\pm$3.5	&		\\[1mm]
2001	May		&	\xmm	&	2.4$\pm$0.3	&	55\err{7}{2}	&	2.2$\pm$0.3	&	$\approx$10	\\[1mm]
2003	 Jun		&	\xte		&	6.5$\pm$0.6	&	- 	&	3.6$\pm$0.8	&	36$\pm$10	\\[1mm]
2003	 Oct		&	\xte		&	13.2$\pm$0.2	&	15$\pm$2	&	3.6*	&	36*	\\[1mm]
2004	 Feb		&	\xte		&	6.5$\pm$0.2	&	-	&	3.6*	&	36*	\\[1mm]
2004	 Aug		&	\xte		&	8.8$\pm$0.2	&	16$\pm$5	&	3.6*	&	36*	\\[1mm]
2005 Apr		&	\xmm	&	4.0$\pm$0.4	&	130\err{6}{7}	&	2.3$\pm$0.1	&	9.7\err{0.8}{1.8}	\\[1mm]
2007 Apr 		&	\xmm	&	7.6$\pm$0.4	&	33\err{4}{5}	&	2.4$\pm$0.9	&	9.3$\pm$2.1	\\[1mm]
2007 May 		&	\suzaku	&	5.3$\pm$0.3	&	44\err{3}{2}	&	2.5$\pm$0.5	&	9.3*	\\[1mm]
2007 May		&	\suzaku	&	4.1$\pm$0.3	&	68\err{6}{7}	&	2.2$\pm$0.4	&	9.3*	\\[1mm]
2007 Nov 		&	\suzaku	&	3.2$\pm$0.2	&	110\err{11}{14}	&	2.2$\pm$0.4	&	9.3*	\\[1mm]
2007	 Nov		&	\suzaku	&	2.6$\pm$0.5	&	120$\pm$20	&	2.4$\pm$0.3	&	9.3*	\\[1mm]
2012	 Aug 		&	\nustar	&	4.8$\pm$0.1	&	24\err{3}{7}	&	2.1$\pm$0.4	&	10$\pm$3	\\[1mm]
2012	 Sep		&	\nustar	&	3.2$\pm$0.1	&	56\err{10}{20}	&	2.1*	&	10*	\\[1mm]

\enddata
\tablecomments{Comparing observed flux, absorption, and reflection measurements over time with the traditional \pexrav model.  A  star (*) indicates a tied or frozen parameter.  A dash (-) indicates an unconstrained parameter.  Early models did not include the reflection hump, which may affect measured \nh and Fe line flux values.  References: Risaliti \etal 2002 (and references therein), T00, P07, B09, and this work. Note that analysis of the four epochs of \xte data are presented in Appendix A.}
\tablenotetext{A}{Flux is in units of $10^{-12}$~\fluxunits}
\tablenotetext{B}{Fe line flux is in units of $10^{-5}$~\feunits.}
\tablenotetext{C}{\pexrav Norm is in units of $10^{-3}$~\crhunits.}
\end{deluxetable}

\section{Discussion and Conclusions}

Both the pure reflection (\pexrav) and obscuring torus (\mytorus) models fit our data equally well. 
When we applied the pure reflection model to our \nustar data we found that the strength of the reflection hump was much higher than is typically predicted by reflection models.
This could be explained by a partial ``shut off,'' i.e., a long-term reduction in flux that started sometime between the 
1998 \sax observation and the 2001 \xmm observation, as suggested by P07, and that the reflection has not yet caught up.
Alternatively, if the torus is intersecting the line of sight, then the obscured continuum component from the ``hidden nucleus''  provides the spectral curvature.

B09 argued that the spectral variability of this source (and indeed many Seyferts) could be attributed to variable obscuration.
Both our models require highly variable moderate levels of absorption, as attributed to the BLR by B09 who found variations on timescales as short as 20 hours.
In order to fully account for the spectral curvature their models also required moderate levels of reflection.
We now have more sophisticated models that allow us to self-consistently model Compton-thick absorption and Compton scattering
from the obscuring torus to better evaluate the hidden nucleus scenario. 
Additionally, \nustar provides the highest quality data to date in the 3--80 keV bandpass.
Figure \ref{figall} shows a simultaneous fit with \nustar and \suzaku data, demonstrating our improved ability
to quantify the amount of reflection in this source and to test the new physical models.

\subsection{The Shut-Off Scenario}

It has been 16 years since the bright \sax observation in 1998 (T00) and the reflection signatures 
have not shown the substantial reduction predicted by P07 (see Figure \ref{figlc} and Table \ref{tabhist}). 
With no X-ray data between 1998 and 2000 this means a lower limit of 12 years until the 2012 \nustar observations,
implying a minimum inner reflecting radius of the material at 6 light years (2~pc) from the source.  
We note that \textsl{HST} data put an upper limit on the size of the nuclear region of 8~pc (Bianchi \etal 2007).

From the unabsorbed X-ray luminosity (measured in 1998) of $\sim 7 \times 10^{41}$ erg s\e{-1} we can estimate a bolometric luminosity of 
$\sim 2 \times 10^{43}$ erg s\e{-1} (see, e.g., Vasudevan \etal 2010) corresponding to a dust sublimation radius of $\sim$0.06~pc.
This is consistent with mid-infrared interferometry results for Seyfert tori that predict a torus size of 0.1--0.5~pc for this luminosity (Burtscher \etal 2013, their Figure 36), 
with the hot inner wall predicted from near-infrared dust reverberation at 0.01--0.1~pc (e.g., Suganuma \etal 2006).
An estimation of the torus size from fitting a clumpy torus model to infrared data gives an approximation of the median torus size of 1.5 pc (Alonso-Herrero \etal 2011).
This is in conflict with the shut-off theory as it makes it unlikely that the torus is larger than the necessary 2~pc.
It is possible that the bulk of the reflection signal is originating outside the torus, for instance in the extended dust lanes of the galaxy,
however the covering factor of the dust lane would likely be far too low to explain the extreme strength of the reflection and is optically thin in the line of sight.   

Other observations between the mid 1990's and today also detract from the shut-off interpretation.
\textsl{HST} observed the source twice, in 1995 June and 2001 July, and found that the optical source flux had increased by a factor of 60\% 
(Bianchi \etal 2007), inconsistent with the source shutting off during this time.
Additionally, \xte and \bat have monitored the source from 2003 to 2011 and found a highly variable observed flux, 
the brightest of which was nearly as bright as the \sax observation, though relatively short lived.
Soldi \etal (2014) performed variability analysis on the \bat light curves of 110 Seyferts, including NGC~7582.
Their results show that above 35 keV, the variability estimator of NGC~7582 is consistent with that of Seyferts in general, 
which rules out a dominant contribution to the high energy flux from reflection.

\subsection{The Hidden Nucleus}

With the advent of physical torus models we are able to investigate fully an alternative to this scenario: that the source has remained at a relatively 
stable average X-ray luminosity (though with high intrinsic source variability) while increased obscuration by the torus has weakened the observed signal in the past decade.  
Our obscuring torus model shows that this scenario is entirely plausible with a patchy absorber (\nh$\,\sim\,10^{24}$~\colunits) providing a strong reflection signal and heavily absorbed 
primary continuum while allowing for leaked emission of the primary continuum.  Another layer of absorption (\nh$\,\sim\,10^{23}$~\colunits) fully covers this leaked emission.
This scenario is illustrated in Figure \ref{figschema}.

B09 noted that the full-covering absorber accounted for nearly all the source variability observed by \xmm and \suzaku.  
They theorized that this layer might be associated with hidden BLR clouds close to the central source 
due to the rapidity of the variations. We confirm this variability with a $50\%$ change in the absorption column in 15 days.  
We also see some variability in the normalization of this leaked power law, a drop of $30\%$ 
which could be explained by an increase in the covering fraction of the torus clouds in the line of sight.

We found in our application of the \pexrav model that the Fe \ka line strength is unexpectedly weak.  
The Fe line strength is about half what we would have expected were it to arise in the same material as the Compton reflection hump.
This could be explained by a low Fe abundance ($\sim 0.6$) or obscuration of the Fe line region by Compton thin material.
The \mytorus model does not find this mismatch, the Fe line strength matching well with the modeled torus.

The torus model implies a much higher intrinsic luminosity for the mostly obscured X-ray emitting corona.
The unabsorbed 2--10 keV fluxes from Table \ref{tabpar} correspond to intrinsic luminosities of $L_{2-10}=$5.3 (4.2) \ten{41} erg s\e{-1} for observations 1(2)
from the \pexrav model and $L_{2-10}=$3.5 \ten{41} erg s\e{-1} for the \mytorus model.

The infrared 6$\mu$m flux of the source is 0.183 Jy (Lutz \etal 2004) corresponding to an infrared luminosity of 5\ten{42} erg s\e{-1}. 
Using empirical relations we predict an intrinsic 2--10 keV X-ray flux of $\sim$5--10 \ten{-11} \fluxunits (e.g., Gandhi \etal 2009, Stern 2015), 
which is consistent with the unabsorbed source flux in our hidden nucleus model, 6  \ten{-11} \fluxunits. 

If we accept the torus model as the most plausible physical interpretation of the data, we find that this source includes three distinct absorbers: 
the slowly varying patchy torus ($3.6 \times10^{24}$~\colunits), the rapidly varying full covering absorber ($\sim\,3-12 \times 10^{23}$~\colunits), and the dust lane 
in the host galaxy ($\sim\,10^{22}$~\colunits), seen by \chandra and included in our models, though not directly measured due to its degeneracy with the full-covering absorber.
 The Compton-thick absorbed continuum is degenerate with the reflection hump in spectral modeling, however it should not affect the measurements of the 
$10^{23}$~\colunits absorption that have been performed by, e.g., B09 and others.  We might expect similar variability in Compton-thick absorption from the patchy torus on longer timescales.


\subsection{Summary}

From our spectroscopic analysis of NGC~7582 we find that it is plausible that this source
could indeed contain a hidden nucleus absorbed by a patchy torus intersecting the line of sight with a covering fraction of $\sim\,80-90\%$.
The more traditional phenomenological model that has been used to describe this source in many recent papers 
fits the data equally well, however, the \nustar data require a higher than expected reflection fraction.
One suggested physical explanation, that the source has switched off, is not viable since it requires a torus at a distance of $\gtrsim$2~pc,
contrary to predictions from infrared reverberation and interferometry measurements of AGN.

Much of the variability of the source is also consistent with patchy absorption.  Continuum flux changes could be due to
a combination of a changing covering factor of Compton-thick material in the line of sight and intrinsic luminosity changes.
Short term spectral variation observed in this source over the past decade has been attributed to changing column density in the 
full covering Compton-thin ($\sim\,10^{23}$~\colunits) BLR clouds as seen by B09.  Additional full covering absorption ($\sim\,10^{22}$~\colunits) 
from the galactic scale dust lane has also been observed.  Detection of the extended soft X-ray emitting gas is consistent
with previous observations by \chandra and \xmm.

\begin{acknowledgments}

We would like to thank the anonymous referee for helpful and constructive comments contributing to this work.
This work was supported under NASA Contract No. NNG08FD60C and sub-contract No. 44A-1092750. 
We have made use of data from the \nustar mission, a project led by
the California Institute of Technology, managed by the Jet Propulsion
Laboratory, and funded by the National Aeronautics and Space
Administration.  We thank the \nustar Operations, Software and
Calibration teams for support with the execution and analysis of
these observations.  This research has made use of the \nustar
Data Analysis Software (NuSTARDAS) jointly developed by the ASI
Science Data Center (ASDC, Italy) and the California Institute of Technology (USA).
This work has made use of also made use of \xte and \swift archival data and 
HEASARC online services, supported by NASA/GSFC, and the NASA/IPAC 
Extragalactic Database, operated by JPL/California Institute of Technology under contract with NASA.
M.B. acknowledges support from NASA Headquarters under the NASA Earth and Space Science Fellowship Program, grant NNX14AQ07H.
P.A. acknowledges financial support form Fondecyt grant 1140304.
F.E.B. acknowledges support from CONICYT-Chile (Basal-CATA PFB-06/2007, FONDECYT 1141218, ``EMBIGGEN" Anillo ACT1101), and the Ministry of Economy, Development, and Tourism's Millennium Science Initiative through grant IC120009, awarded to The Millennium Institute of Astrophysics, MAS.
M.K. acknowledges support from the Swiss National Science Foundation and Ambizione fellowship grant PZ00P2\textunderscore154799/1.

\end{acknowledgments}

{\it Facilities:} \facility{NuSTAR}, \facility{Swift}, \facility{RXTE}


\begin{appendix}

\subsection{A. \xte Reduction and Analysis}

\xte observed NGC~7582 from 2003 to 2004 with four clusters of 3 to 10 exposures each.
The light curve and stacked spectrum for these observations were published in Rivers \etal (2011) and Rivers \etal (2013),
with the caveat that the source was highly variable over these observations on timescales as short as a few days, both in flux and hardness.  
In order to investigate this variability we have grouped the spectra by month: 
2003 June (10.3~ks total net exposure), 2003 October (83.6~ks), 2004 February (45.4~ks) and 2004 August (45.7~ks).
The PCA data were extracted using HEASOFT version 6.7 software (no updates to \xte extraction procedures have been included since).
We extracted PCA STANDARD-2 data from PCU 2 using only events from the top Xe layer in order to maximize signal-to-noise.
Standard screening was applied with time since SAA passage $>$20 minutes and background model file ``pca\_bkgd\_cmfaintl7\_eMv20111129.mdl."

We performed spectral fitting on all four epochs of \xte data between 3.5 and $\sim$ 20 keV.
Naive modeling of these data is difficult owing to the low spectral quality.  
We therefore applied both the \pexrav and \mytorus models to the four \xte epochs simultaneously,
omitting the soft power law and lines and tying parameters in order to reduce the number of free parameters.

\begin{figure}[h]
   \includegraphics[width=0.45\textwidth]{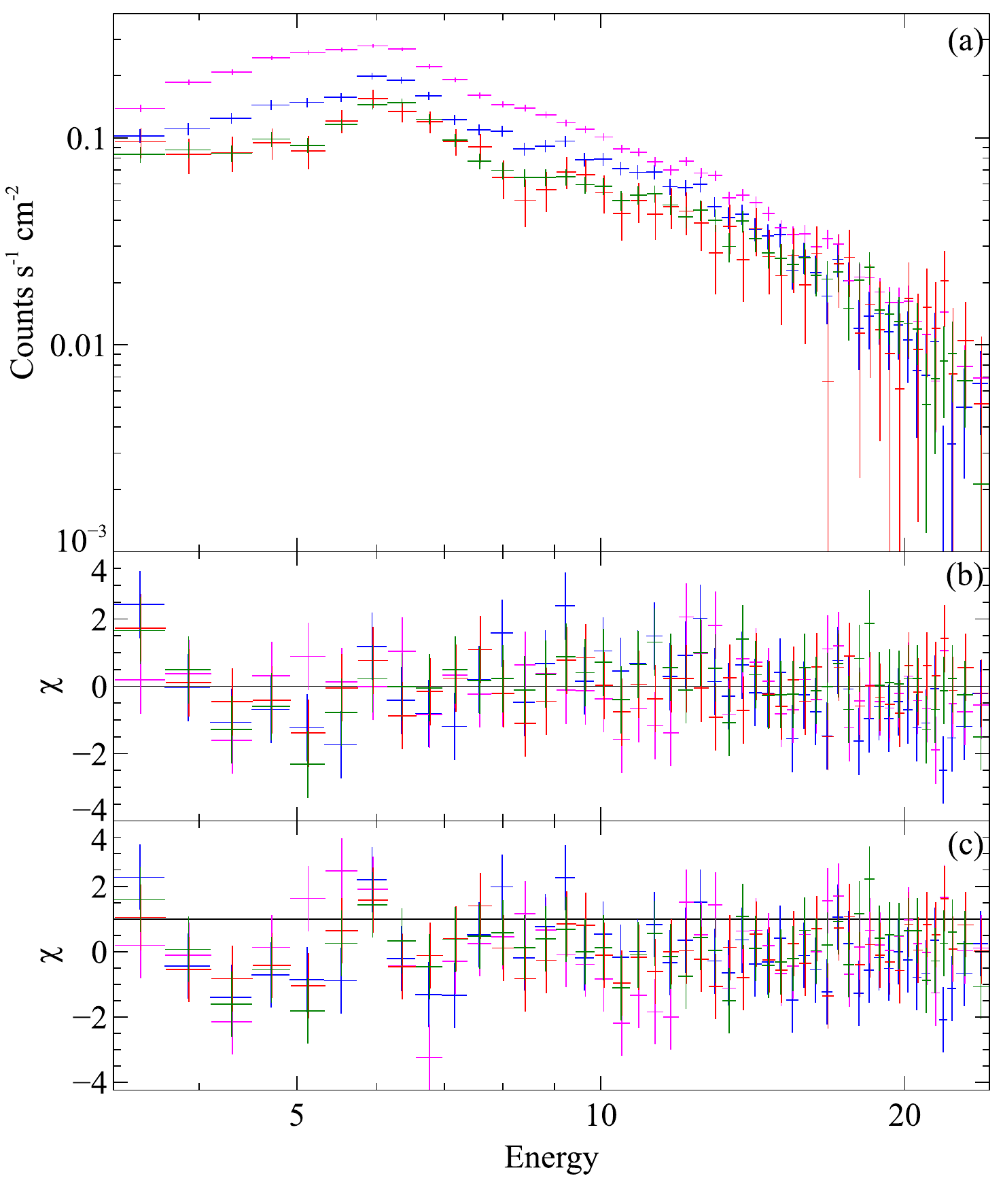}
 \caption{\xte spectrum and residuals to the \pexrav and \mytorus models. Red data are from 2003 June, magenta data are from 2003 October, blue data are from 2004 February, green data are from 2004 August.  a) \xte data from four epochs.  b) Data/model ratios to \pexrav best fit model. c)  Data/model ratios to \mytorus best fit model.}
  \label{figrxte}
\end{figure}

For the \pexrav model we initially tied the photon index and Fe line energy (width frozen) between the spectra.
The Fe line and \pexrav normalizations were consistent across all four epochs and we tied those parameters as well.
Best fit parameters are listed in Table \ref{tabxte} and data and data/model ratio plots are shown in Figure \ref{figrxte}.
The first epoch was highly reflection dominated with a weak continuum which did not allow us to place constraints on the absorption.
Epochs two and four showed very similar levels of absorption (\nh$ \sim 2 \times 10^{23}$~\colunits) while
epoch three appeared to be heavily absorbed (\nh$ = 1.4 \times 10^{24}$~\colunits) and reflection dominated.
There is strong degeneracy between the \pexrav strength and photon index in these data.  If we freeze the photon 
index to 1.9 (1.8) to bring it more in line with other measurements, we find that the \pexrav strength drops to $\sim$25 (15) $\times 10^{-3}$~\crhunits,

Applying the \mytorus model to these data, we found strong consistency between the epochs with a torus column 
density of \nh$ = (2.5 \pm 0.3) \times 10^{24}$~\colunits, $\Gamma=2.1$, and \chidof=156/181.
The only parameters required to be left free were the full covering absorption and leaked power law normalization.  
The leaked power law portions were $6\pm1$\%, $20\pm3$\%, $7\pm1$\%, and $15\pm2$\% for each epoch, respectively.
Allowing the Fe line energy to be free also led to an improvement in the fit (\chidof=143/177) and a visual improvement 
in the residuals around 5--6 keV, however is likely not real but calibration related.

\begin{deluxetable*}{lcccccccr}
   \tablecaption{Broadband Model Parameters \label{tabxte}}
   \tablecolumns{9}
   \startdata
\hline
\hline\\[-1mm]
\pexrav & Observed		& Unabsorbed			 	& Photon  	&	PL		& 	Column  		& 	Fe \ka	& \pexrav 			&	$\chi^2$/dof \\[1mm]
Model  & F$_{2-10}$\tablenotemark{A}	& Power Law  	& Index 		&	Norm\tablenotemark{B}	& 	Density \tablenotemark{C}		& Line Norm\tablenotemark{D} &	Norm\tablenotemark{E}	 \\[1mm]
	   &  		& F$_{2-10}$\tablenotemark{A} 		& ($\Gamma$) 	&  &  (\nh) & 	&	 \\[1mm]
\hline\\
\xte 03-Jun 	&	6.5$\,\pm\,$0.6		&	10$\,\pm\,$8	&	2.1$\,\pm\,$0.1	&	0.4$\,\pm\,$0.4	&	\textsl{unconstr.	} &	3.6$\,\pm\,$0.8	&	36$\,\pm\,$10	&	157/181	\\[1mm]
\xte 03-Oct		&	13.2$\,\pm\,$0.2		&	17$\,\pm\,$2	&				&	7.2$\,\pm\,$1.0	&	15$\,\pm\,$2	&			\\[1mm]
\xte 04-Feb 	&	6.5$\,\pm\,$0.2	&	11$\,\pm\,$7	&				&	0.5$\,\pm\,$0.3	&	\textsl{unconstr.	}	&			\\[1mm]
\xte 04-Aug 	&	8.8$\,\pm\,$0.2		&	8$\,\pm\,$2	&				&	3.1$\,\pm\,$0.6	&	16$\,\pm\,$5	&			\\[1mm]
\hline
\hline\\[-1mm]
\mytorus	 & Observed		& Unabsorbed 			& Photon  				& Absorbed PL			&	Leaked PL  		& Full Covering  			&  Torus Column 		&  $\chi^2$/dof \\[1mm]
Model	& F$_{2-10}$\tablenotemark{A}	 & Power Law  	& Index 				& Norm\tablenotemark{B}				&	Norm\tablenotemark{B}			& Column Density\tablenotemark{C}  		& Density\tablenotemark{C}			&  \\[1mm]
		& &  F$_{2-10}$\tablenotemark{A}			& ($\Gamma$) 			&		&	 		& (\nh) & (\nh)	&    \\[1mm]
\hline\\
\xte 03-Jun 	&	6.5$\,\pm\,$0.6	& 100$\,\pm\,$20 & 2.1$\,\pm\,$0.1 &	44$\,\pm\,$10	&	3.0$\,\pm\,$0.7	&	7$\,\pm\,$5	&	250$\,\pm\,$30	&	166/181	\\[1mm]
\xte 03-Oct	 	&	13.1$\,\pm\,$0.3	&	&			&				&	13$\,\pm\,$2	&	15$\,\pm\,$1		\\[1mm]
\xte 04-Feb 	&	6.5$\,\pm\,$0.2	&	&			&				&	3.5$\,\pm\,$0.6	&	10$\,\pm\,$4		\\[1mm]
\xte 04-Aug 	&	8.7$\,\pm\,$0.2	&	&			&				&	7.1$\,\pm\,$0.9	&	15$\,\pm\,$2		\\[-1mm]
\enddata
\tablecomments{Model parameters to simultaneous fitting of the four epochs of \xte spectral data with the \pexrav and \mytorus models.  
Note we have fixed the inclination angle to 65\degr for both models.}
\tablenotetext{A}{Flux is in units of $10^{-12}$~\fluxunits.}
\tablenotetext{B}{Power law norm is in units of $10^{-3}$~counts s\e{-1} keV\e{-1} at 1 keV.}
\tablenotetext{C}{Column densities are in units of $10^{22}$\,cm$^{-2}$}
\tablenotetext{D}{Fe line flux is in units of $10^{-5}$~\feunits.}
\tablenotetext{E}{\pexrav Norm is in units of $10^{-3}$~\crhunits.}
\end{deluxetable*}
\end{appendix}


\end{document}